\documentclass[floatfix,a4paper,tightenlines,amsmath,amssymb,nofootinbib,showpacs,notitlepage,showkeys,prd,twocolumn,10pt,aps,superscriptaddress]{revtex4-1}

\usepackage[utf8]{inputenc}
\usepackage{amsmath}
\usepackage{amssymb}
\usepackage{graphicx}
\usepackage{hyperref}
\usepackage[export]{adjustbox}
\usepackage[usenames,dvipsnames,svgnames,table]{xcolor}

\newcommand{\be}{\begin{eqnarray}}
\newcommand{\ee}{\end{eqnarray}}

\newcommand{\ba}{\begin{array}}
\newcommand{\ea}{\end{array}}
\newcommand{\bee}{\begin{equation}\ba{c}}
\newcommand{\eee}{\ea\end{equation}}

\newcommand{\bi}{\begin{itemize}}
\newcommand{\ei}{\end{itemize}}

\newcommand{\lsim}{{\;\raise0.3ex\hbox{$<$\kern-0.75em\raise-1.1ex\hbox{$\sim$}}\;}}
\newcommand{\gsim}{{\;\raise0.3ex\hbox{$>$\kern-0.75em\raise-1.1ex\hbox{$\sim$}}\;}}

\newcommand{\beq}{\begin{equation}}
\newcommand{\eeq}{\end{equation}}
\newcommand{\bea}{\begin{eqnarray}}
\newcommand{\eea}{\end{eqnarray}}

\newcommand{\ra}{\rightarrow}

\begin{document}

\title{Comments on  bosonic super-WIMPs search experiments}

\newcommand{\cupibs}{\affiliation{Center for Underground Physics, Institute for Basic Science (IBS), Daejeon 34126, Republic of Korea}}
\newcommand{\ku}{\affiliation{Department of Accelerator Science, Korea University, Sejong 30019, Republic of Korea}}

\author{Young Ju Ko}
\email{yjko@ibs.re.kr}
\affiliation{Center for Underground Physics, Institute for Basic Science (IBS), Daejeon 34126, Republic of Korea}
\author{HyangKyu Park}
\email{Correspoinding author : hyangkyu@korea.ac.kr}
\affiliation{Department of Accelerator Science, Korea University, Sejong 30019, Republic of Korea}

\date{\today}

\begin{abstract}
Bosonic super-WIMPs, including pseudoscalar and vector particles, are dark matter candidates. 
Until now, many underground experiments searches for super-WIMPs have been performed 
in the mass range of a few  $\rm keV/c^2$ to 1 $\rm MeV/c^2$. 
All these searches utilize the absorption process  of a super-WIMP 
by a target atom in the detector, which is similar to the photoelectric effect.
In this study, we consider another process, namely, a Compton-like process.
As an example, we compare the cross-section of a germanium atom for 
the absorption process with that of a Compton-like process.
Our findings indicate that the cross-section for the Compton-like process becomes dominant, 
compared to that for the absorption process
for mass above approximately 150 $\rm keV/c^2$  for both  pseudoscalar and vector super-WIMPs.
In particular, the cross-section for the Compton-like process for a vector super-WIMP becomes 
increasingly greater than that for the absorption process
by an order of magnitude to two orders of magnitude in the 400 $\rm keV/c^2$ to 1 $\rm MeV/c^2$ mass range, respectively.
By including the Compton-like process, which has not been used in any other super-WIMP search experiment, the
experimental upper limits can be improved.
\end{abstract}



\maketitle

\section{Introduction}
One of the most fundamental questions in physics is the nature of dark matter (DM). 
The existence of DM, which constitutes approximately 80\% of the matter in the universe~\cite{Ade:2015ira}, 
has been inferred from a wide range of astrophysical and cosmological systems 
based solely on the gravitational effects~\cite{Bergstrom:2012fi, Feng:2010gw}. 
Although the standard model (SM) of particle physics is extraordinarily successful, it cannot explain the nature of DM. 

The freeze-out mechanism is a plausible method for the generation of the DM population in our universe. 
 It has been assumed that DM particles, commonly called weakly interacting massive particles (WIMPs), 
 interacted with SM particles in thermal equilibrium at very high temperatures 
 in the early universe, and the DM particles that were out of equilibrium remained as the universe expanded. 
 The currently observed dark-matter relic abundance has been generated 
 due to the weak mass scale $\mathcal{O}(\rm 100~GeV)$ of DM with
 weak interaction strength between the WIMPs and SM particles~\cite{Lee:1977ua,Vysotsky:1977pe}.
 The DM produced by this mechanism is commonly referred to as a WIMP. 
Extensive searches over the past 40 years for a WIMP with $\mathcal{O}(\rm 100~GeV)$ mass 
have not yielded any result.
Hence, it is desirable to investigate various theoretical models of DM.
 
Based on the numerical simulation results for the small-scale structure formation of a WIMP based on gravitational 
 interaction, the DM mass is considered to be greater than $\sim$~3 keV~\cite{Markovic:2013iza}. 
 A mass on this scale implies weak interaction between the DM and SM particles. 
Several models have been proposed with DM mass scales ranging from keV to MeV, 
for the so-called super-WIMP~\cite{Pospelov:2008jk,Pospelov:2007mp,Dodelson:1993je, Dolgov:2000ew,Ellis:1984er,Feng:2003uy} .
 Fermionic super-WIMPs, sterile neutrinos, and gravitinos have been thoroughly studied, and are extremely difficult to detect experimentally. 
 On the other hand, bosonic super-WIMPs allow the decay of lighter SM particles~\cite{Pospelov:2008jk}, and such decays with super-weak interactions can be strongly suppressed. 
Nonetheless, super-WIMPs can be absorbed or emitted by SM particles, which can be directly detected through terrestrial experiments. In particular, their absorption process in the target material deposits energy into the target atom, which 
corresponds to the rest mass of the super-WIMP.

Experimental explorations for pseudoscalar (vector) super-WIMPs, known as axions (dark photons), have been performed based on their absorption in detector material. The constraints for the coupling strength of super-WIMPs with electrons for the super-WIMP mass have been determined through
underground experiments~\cite{Abe:2018owy, Fu:2017lfc,Akerib:2017uem,Aprile:2017lqx,GERDA:2020emj,Aralis:2019nfa,Armengaud:2018cuy,Liu:2016osd,Abgrall:2016tnn, Angloher:2016rji,Adhikari:2019tgv},
which investigate the super-WIMP mass region from $\sim \rm keV/c^2$ to 1 $\rm MeV/c^2$.
To detect super-WIMP events, all these studies utilize the absorption process of the super-WIMP by target atoms,
which is similar to the photoelectric effect; the cross-sections for this process are proportional to those of the photoelectric
effect, where the photon energy is equal to the super-WIMP mass. 
When using the absorption process for super-WIMP detection, the upper limit becomes weak with the increase in the super-WIMP mass. 
For example, the cross-section of Ge for the photoelectric effect 
significantly decreases from 4,306 barn/atom at 10 keV to 0.0086 barn/atom at 1 MeV.

In this study, we consider a Compton-like process 
$a + e^- \ra e^- + \gamma$ ($V + e^- \ra e^- + \gamma$) for a pseudoscalar (vector) super-WIMP, 
where $a$ and $V$ represent a pseudoscalar and vector super-WIMP, respectively.
In general, the cross-section for the Compton process is greater than that for the photoelectric effect 
at photon energies above $\sim \rm 100~keV$.
By including the Compton-like process, we expect an improvement in the experimental sensitivity 
in the mass range of $\sim \rm 100~keV/c^2$ to a few $\rm MeV/c^2$.  
The most recent search for super-WIMP used Ge atoms as the target material
in the GERDA experiment~\cite{GERDA:2020emj} for a super-WIMP mass range up to 1 $\rm {MeV/c^2}$.
We study the cross-section for the absorption and Compton-like processes with a
Ge target for a super-WIMP model
in the 1 $\rm keV/c^2$ to 1 $\rm MeV/c^2$ mass range.  
 
\section{The model for super-WIMP}
\label{sec:model}
In this section, we discuss the absorption process, which is similar to the photoelectric effect, and a Compton-like process for the interaction of a
 pseudoscalar super-WIMP and vector super-WIMP with an atom and electron, respectively. 
 
The Lagrangian for the interaction of pseudoscalar super-WIMP $a$ with an electron ~\cite{Pospelov:2008jk} can be expressed as follows: 
\bea
\mathcal{L} = 2 \frac{m_e}{f_a} a \bar{\psi} i \gamma_5 \psi \label{eq:lagpsedo},
\eea 
where $f_a$ is the interaction strength, $m_e$ is the electron mass, and $\psi$ is the electron field.
Therefore, this interaction Lagrangian gives the cross-section 
for the absorption process $\sigma_{a,abs}$~\cite{Pospelov:2008jk} with an atom, which is the so-called axioelectric effect, 
in terms of the cross-section for the photoelectric effect $\sigma_{pe}$, with the photon energy $\omega$ equivalent to
the mass of the pseudoscalar super-WIMP ($m_a$):
\bea
\frac{\sigma_{a,abs} v}{\sigma_{pe}(\omega=m_a) \rm c} \approx \frac{3 m_a^2}{4\pi \alpha f_a^2} \label{eq:xsecpsedo},
\eea 
where $v$ is the incoming velocity of the super-WIMP, $\rm c$ is the velocity of light, and $\alpha$ is the fine structure constant.
Eq.~\eqref{eq:xsecpsedo} can be rewritten by introducing dimensionless coupling $g_{aee}=2m_e/f_a$ as follows:
\bea
\frac{\sigma_{a,abs}v}{\sigma_{pe}(\omega=m_a) \rm c} \approx g_{aee}^2 \frac{3 m_a^2}{16 \pi \alpha m_e^2}. \label{eq:xsecpsedo1}
\eea 

For the interaction of vector super-WIMP $V$ with an electron, the Lagrangian~\cite{Pospelov:2008jk} is given by
\bea
\mathcal{L} = e \kappa V_\mu \bar{\psi} \gamma^\mu \psi, \label{eq:lagvector}
\eea  
where $V_\mu$ is the field of the super-WIMP, e is the electron charge, 
and $\kappa$ is the kinetic mixing parameter with the electromagnetic field. 
The absorption cross-section of the vector super-WIMP ($\sigma_{V,abs}$) 
by an atom~\cite{Pospelov:2008jk} is calculated as 
\bea
\frac{\sigma_{V,abs}v}{\sigma_{pe}(\omega=m_V) \rm c} \approx \frac{\alpha'}{\alpha} \label{eq:xsecvec},
\eea
where $m_V$ is the mass of the vector super-WIMP and $\alpha'$ is $\frac{(e\kappa)^2}{4 \pi}$.

The Compton-like process where a pseudoscalar super-WIMP or vector super-WIMP interacts with an electron and produces 
a photon ($\gamma$) is as follows:
 \bea
 e + \phi &\rightarrow& e +\gamma\label{eq:cmpprocess}, 
\eea
where $\phi$ can be either a pseudoscalar super-WIMP  or vector super-WIMP.
The cross-section for the Compton-like process of a pseudoscalar (vector) super-WIMP, 
$\sigma_{a,cmp}$ ($\sigma_{V,cmp}$)~\cite{Liu:2017htz}, can be determined as follows: 
\begin{subequations}
\label{eq:xseccmp}
\bea
\label{eq:xseccmp:1}
\sigma_{a,cmp}&=& g_{aee}^2\frac{\alpha}{4 m_e |\mathbf{k}|}\int_{-1}^1 d\cos\theta_\gamma\frac{|\mathbf{q}|\mathcal{A}_P}{E_k+m_e-|\mathbf{k}|\cos\theta_\gamma},~~~~~ \\
\label{eq:xseccmp:2}
\sigma_{V,cmp}&=&\kappa^2\frac{\pi\alpha^2}{3m_e |\mathbf{k}|}\int_{-1}^1 d\cos\theta_\gamma\frac{|\mathbf{q}|\mathcal{A}_V}{E_k+m_e-|\mathbf{k}|\cos\theta_\gamma},~~~
\eea
\end{subequations}
where $\mathbf{k}$ is the momentum of the incoming super-WIMP, $\mathbf{q}$ is the momentum of the outgoing photon,
$\theta_\gamma$ is the angle between the photon and super-WIMP, and $\mathcal{A}_P$ and $\mathcal{A}_V$ are
 the matrix elements of the Compton-like process 
 for the pseudoscalar super-WIMP and vector super-WIMP~\cite{Liu:2017htz}, respectively.
Here, $|\mathbf{q}|$ for the Compton-like process in \eqref{eq:cmpprocess} is given by:
\bea
 |\mathbf{q}| = \frac{m_\phi^2+ 2E_k m_e }{2(m_e + E_k - |\mathbf{k}| \cos\theta_\gamma)},
\eea 
where $m_\phi$ is the mass of the super-WIMP, and $E_k$ is the energy of the super-WIMP, which can be considered as the mass of the super-WIMP.

\section{Cross-section estimations for the absorption and Compton-like processes}
\label{sec:xsec}
In this section, we discuss the cross-section of the Ge atom for the absorption and Compton-like processes. 

For estimating the cross-section for the absorption process, 
we used Eq.~\eqref{eq:xsecpsedo1} for a pseudoscalar super-WIMP 
and Eq.~\eqref{eq:xsecvec} for a vector super-WIMP.
For such estimation, the cross-sections for the photoelectric effect ($\sigma_{pe}$) of the Ge atoms were used in~\cite{photoelect}. 

For an atom, the cross-sections of the Compton-like process for a pseudoscalar and vector super-WIMP are given by
\begin{subequations}\label{eq:xseccmpatm}
\bea
\label{eq:xseccmpatm:1}
 \sigma_{a,cmp}^{Z} &=& N_e  \sigma_{a, cmp}, \\
\label{eq:xseccmpatm:2} 
 \sigma_{V,cmp}^{Z} &=& N_e  \sigma_{V,cmp},
\eea
\end{subequations}
where $\sigma_{a,cmp}^Z$  ($\sigma_{V,cmp}^Z$) is the cross-section for the Compton-like process 
for a pseudoscalar (vector super-WIMP) for an atom and 
$N_e$ is the number of electrons (atomic number) in the atom.
$\sigma_{a,cmp}$ and $\sigma_{V,cmp}$ can be calculated numerically using Eq.~\eqref{eq:xseccmp:1} and Eq.~\eqref{eq:xseccmp:2}, respectively.
 We set $N_e$ = 32 as the atomic number of the Ge atom. 
 
The total cross-section ($\sigma_{atom}$) for a super-WIMP interacting with an atom, 
including both the absorption and Compton-like processes, can be expressed as
\bea
\sigma_{atom} = \sigma_{a (V),abs} + \sigma_{a (V),cmp}^Z.  
\eea

\begin{figure*}[tb]
\includegraphics[width=0.49\textwidth]{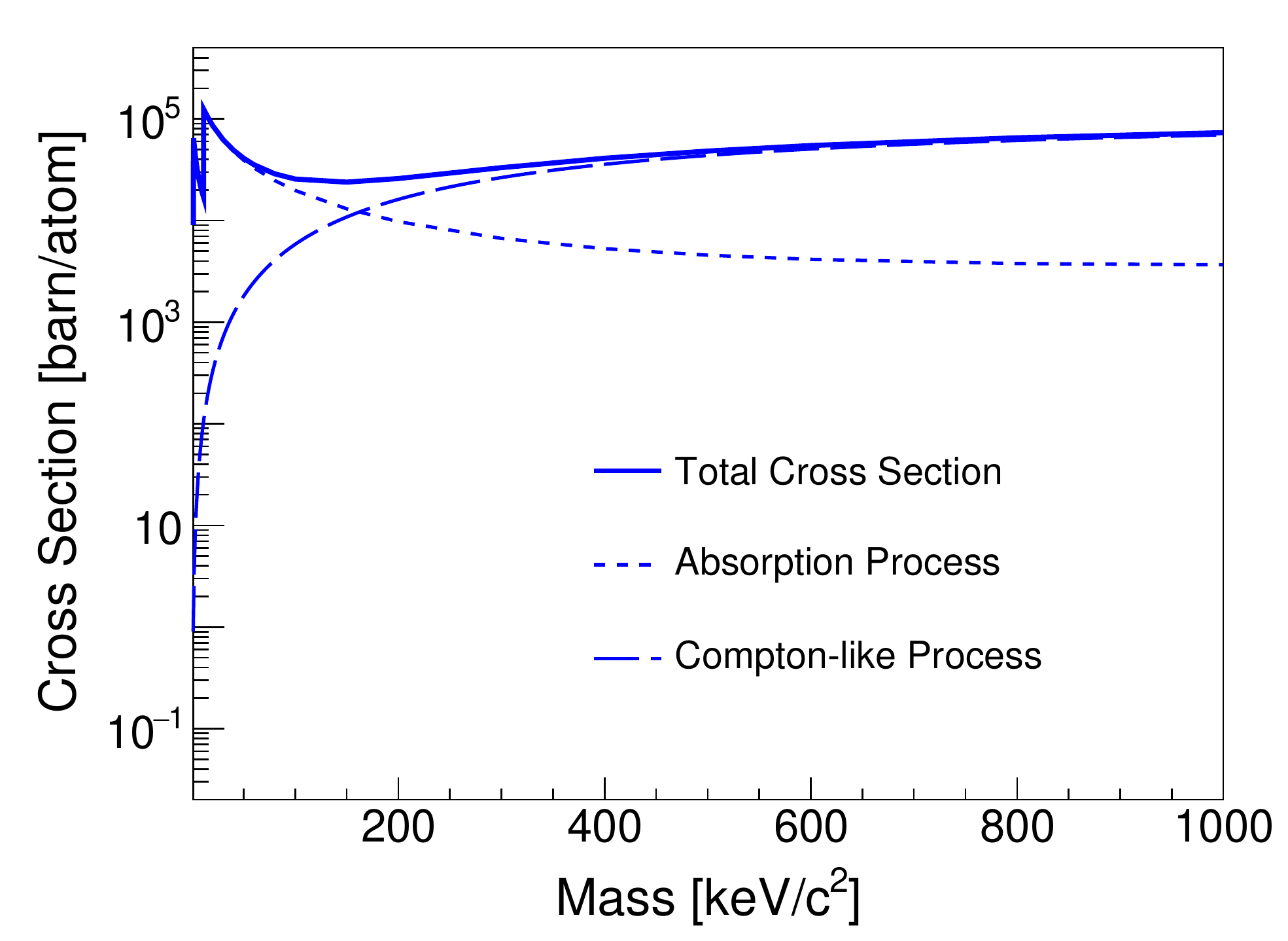}
\includegraphics[width=0.49\textwidth]{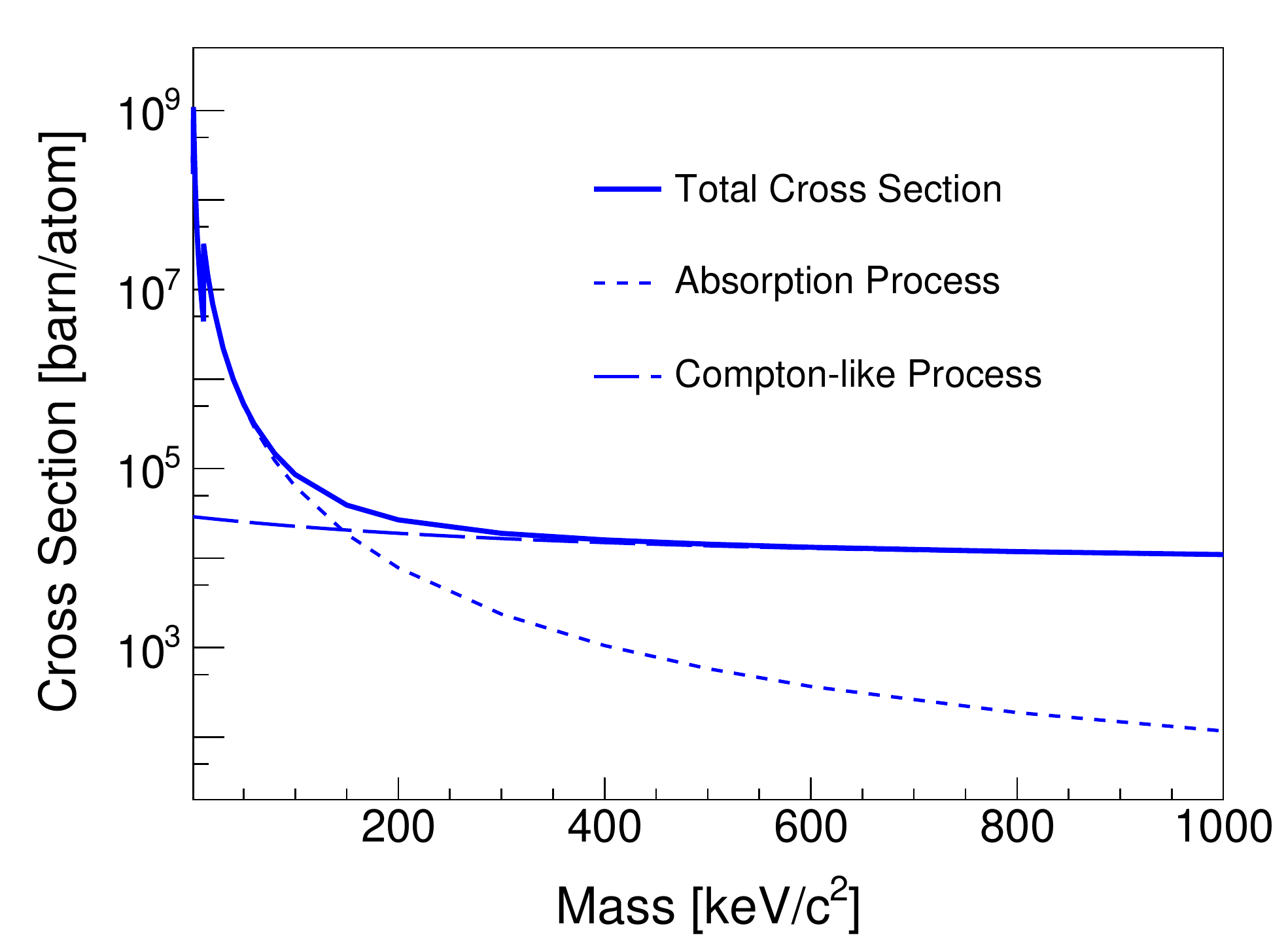}
\caption{Cross-sections (unit: barn/atom) for the absorption and Compton-like processes, 
and the total cross-section, including both processes, 
for the pseudoscalar (left) and vector (right) super-WIMPs as a function of the mass. 
We used the dimensionless coupling constant $g_{aee}=1$ (kinetic mixing parameter $\kappa=1$) 
for a pseudoscalar (vector) super-WIMP.}
\label{fig:xsec}
\end{figure*} 

To compare the cross-sections for the absorption and Compton-like processes for the interaction of a pseudoscalar (vector) super-WIMP
with a Ge atom, we applied the dimensionless coupling constant $g_{aee}=1$ (kinetic mixing parameter $\kappa=1$).
The results for each process and the total cross-section for the pseudoscalar and vector super-WIMP 
are shown in Fig.~\ref{fig:xsec}. 
As previously mentioned, the cross-section for the Compton-like process is greater than that of the absorption process 
in the mass range above approximately 500 $\rm keV/c^2$ (100 $\rm keV/c^2$) for a pseudoscalar (vector) super-WIMP.
In particular, the cross-section for the Compton-like process for a vector super-WIMP continues to increase from
an order of magnitude to two orders of magnitude in the 400 $\rm keV/c^2$ to 1 $\rm MeV/c^2$ mass range, respectively.

\section{Conclusions}
\label{sec:conclusion}
In this study, we examined the absorption and Compton-like processes for the interaction of a super-WIMP with an atom. 
As an example, we compared the cross-sections of both processes for a Ge atom. 
The obtained results demonstrated that the Compton-like process 
had a greater cross section than the absorption process
for mass above approximately 150 $\rm keV/c^2$  for both pseudoscalar  and vector super-WIMPs.
The cross-section for the Compton-like process for a pseudoscalar super-WIMP is increased to 
one order of magnitude at 1 $\rm MeV/c^2$ of mass.
However, the cross-section for the Compton-like process for a vector super-WIMP becomes 
increasingly greater than that for the absorption process
by an order of magnitude to two orders of magnitude in the 400 $\rm keV/c^2$ to 1 $\rm MeV/c^2$ mass range, respectively.
By including the Compton-like process, which has not been used in any other super-WIMP search experiment, the
experimental upper limits can be improved. 
 \\
 \\
 \\
\section*{Acknowledgments}
We acknowledge Ein Park and Euny Lee for their invaluable discussion.
The work of Y.J. Ko was supported by
the Institute for Basic Science (IBS) in Korea under project codes IBS-R016-A1.
HKP was supported by a Korea University Grant and the National Research Foundation of Korea (NRF-2018R1D1A1B07048941),
and the Korea Basic Science Institute (KBSI) National Research Facilities and Equipment Center (NFEC) 
grant funded by the Korean government (Ministry of Education) (No. 2019R1A6C1010027).


\begin{thebibliography}{25}
\bibitem{Ade:2015ira}
P.~A.~R.~Ade \textit{et al.} (Planck Collaboration),
Astron. Astrophys. \textbf{594}, A23 (2016).

\bibitem{Bergstrom:2012fi}
L.~Bergstrom,
Annalen Phys. \textbf{524}, 479 (2012).


\bibitem{Feng:2010gw}
J.~L.~Feng,
Ann. Rev. Astron. Astrophys. \textbf{48}, 495 (2010).



\bibitem{Lee:1977ua}
B.~W.~Lee and S.~Weinberg,
Phys. Rev. Lett. \textbf{39}, 165 (1977).

\bibitem{Vysotsky:1977pe}
M.~I.~Vysotsky, A.~D.~Dolgov and Y.~B.~Zeldovich,
JETP Lett. \textbf{26}, 188 (1977).


\bibitem{Markovic:2013iza}
K.~Markovi\v{c} and M.~Viel,
Publ. Astron. Soc. Austral. \textbf{31}, e006 (2014). 


\bibitem{Pospelov:2008jk}
M.~Pospelov, A.~Ritz and M.~B.~Voloshin,
Phys. Rev. D \textbf{78}, 115012 (2008). 

\bibitem{Pospelov:2007mp}
M.~Pospelov, A.~Ritz and M.~B.~Voloshin,
Phys. Lett. B \textbf{662}, 53 (2008). 

\bibitem{Dodelson:1993je}
S.~Dodelson and L.~M.~Widrow,
Phys. Rev. Lett. \textbf{72}, 17 (1994).


\bibitem{Dolgov:2000ew}
A.~D.~Dolgov and S.~H.~Hansen,
Astropart. Phys. \textbf{16}, 339 (2002).


\bibitem{Ellis:1984er}
J.~R.~Ellis, D.~V.~Nanopoulos and S.~Sarkar,
Nucl. Phys. B \textbf{259}, 175 (1985).

\bibitem{Feng:2003uy}
J.~L.~Feng, A.~Rajaraman and F.~Takayama,
Phys. Rev. D \textbf{68}, 063504 (2003).

\bibitem{Abe:2018owy}
K.~Abe \textit{et al.} (XMASS Collaboration),
Phys. Lett. B \textbf{787}, 153 (2018).

\bibitem{Fu:2017lfc}
C.~Fu \textit{et al.} (PandaX Collaboration),
Phys. Rev. Lett. \textbf{119},  181806 (2017).

\bibitem{Akerib:2017uem}
D.~S.~Akerib \textit{et al.} (LUX Collaboration),
Phys. Rev. Lett. \textbf{118}, 261301 (2017).

\bibitem{Aprile:2017lqx}
E.~Aprile \textit{et al.} (XENON100 Collaboration),
Phys. Rev. D \textbf{96}, 122002 (2017).

\bibitem{GERDA:2020emj}
M.~Agostini \textit{et al.} (GERDA Collaboration),
Phys. Rev. Lett. \textbf{125}, 011801 (2020).

\bibitem{Aralis:2019nfa}
T.~Aralis \textit{et al.} (SuperCDMS Collaboration),
Phys. Rev. D \textbf{101}, 052008 (2020), 
[erratum: Phys. Rev. D \textbf{103} (2021) no.3, 039901].

\bibitem{Armengaud:2018cuy}
E.~Armengaud \textit{et al.} (EDELWEISS Collaboration),
Phys. Rev. D \textbf{98}, 082004 (2018).


\bibitem{Liu:2016osd}
S.~K.~Liu \textit{et al.} (CDEX Collaboration),
Phys. Rev. D \textbf{95}, 052006 (2017).

\bibitem{Abgrall:2016tnn}
N.~Abgrall \textit{et al.} (Majorana Collaboration),
Phys. Rev. Lett. \textbf{118}, 161801 (2017).

\bibitem{Angloher:2016rji}
G.~Angloher \textit{et al.} (CRESST Collaboration),
Eur. Phys. J. C \textbf{77}, 299 (2017).

\bibitem{Adhikari:2019tgv}
P.~Adhikari \textit{et al.} (COSINE-100 Collaboration),
Astropart. Phys. \textbf{114}, 101 (2020).

\bibitem{Liu:2017htz}
Y.~S.~Liu and G.~A.~Miller,
Phys. Rev. D \textbf{96}, 016004 (2017).

\bibitem{photoelect}
Database~webpage, \\
http://physics.nist.gov/PhysRefData/Xcom/html/xcom1.html


\end{thebibliography}

\end{document}